\documentclass{article}
\usepackage{ltwol2e}

\usepackage{epsfig}

\def\Journal#1#2#3#4{{#1}{ \bf #2}, #3 (#4)}

\def\NIMA{{\em Nucl. Instrum. Methods} A}
\def\NPB{{\em Nucl. Phys.} B}
\def\PLB{{\em Phys. Lett.}  B}
\def\PRL{\em Phys. Rev. Lett.}
\def\PRD{{\em Phys. Rev.} D}
\def\ZPC{{\em Z. Phys.} C}
\def\EPC{{\em Eur. Phys. J.} C}
\def\CPC{\em Comp. Phys. Comm.}

\def\be{\begin{equation}}
\def\ee{\end{equation}}
\def\bea{\begin{eqnarray}}
\def\eea{\end{eqnarray}}

\def\sigepres{\sigma^{ep}_{\rm res}}
\def\sigresses{\sigma^{\rm SES}_{\rm res}}
\def\xgam{x_{\gamma}}
\def\xgamjets{x_{\gamma}^{\rm jets}}
\def\xgamrec{x_{\gamma}^{\rm rec}}
\def\xp{x_{p}}

\newcommand{\etajet}[1][]{\ensuremath{\eta_{\mathrm{jet}{#1}}}}

\def\meta{\bar{\eta}}

\def\qqbar{\mbox{q}\overline{\mbox{q}}}
\def\ppbar{\overline{\mbox p}\mbox{p}}
\newcommand{\etjet}[1][]{\ensuremath{E_{\mathrm{T,jet}{#1}}}}
\def\etjetq{E^2_{\rm T, jet}}
\def\et{E_{\rm T}}
\def\etq{E^2_{\rm T}}
\def\pt{p_{\rm T}}
\def\ptq{p^2_{\rm T}}
\def\pthadq{p^2_{\rm T,had}}
\def\ptjet{p_{\rm T,jet}}

\def\met{\bar{E}_{\rm t}}
\def\metq{\bar{E}^2_{\rm t}}
\def\egam{e\gamma}
\def\gamgam{\gamma\gamma}
\def\gamstarp{\gamma^*p}
\def\gamp{\gamma p}
\def\epem{e^+e^-}
\def\ejet{E_{\rm jet}}
\def\eh{E_{\rm h}}
\def\pzjet{p_{z,\rm jet}}
\def\pzh{p_{z,\rm h}}
\def\fgame{f_{\gamma /e}}
\def\figam{f_{i/ \gamma}}
\def\fqgam{f_{q/ \gamma}}
\def\fggam{f_{g/ \gamma}}
\def\fqgamp{f_{q/ \gamma ,p}}
\def\fqbargamp{f_{\bar{q}/ \gamma ,p}}
\def\fggamp{f_{g/ \gamma ,p}}
\def\fjp{f_{j/p}}
\def\mij2{|M_{ij}|^2}
\def\gev2{GeV$^2$}
\def\feffgam{\tilde{f}_{\gamma}}
\def\feffp{\tilde{f}_{p}}
\def\feffgamp{\tilde{f}_{\gamma ,p}}

\newcommand {\gapprox}{\raisebox{-0.7ex}{$\stackrel {\textstyle>}{\sim}$}}

\begin{document}

\setcounter{footnote}{0}
\renewcommand{\thefootnote}{\fnsymbol{footnote}}

\title{PHOTON STRUCTURE
\footnotemark[1]}

\author{G\"UNTER GRINDHAMMER}

\address{Max-Planck-Institut f\"ur Physik, (Werner-Heisenberg-Institut), 
D-80805 M\"unchen, Germany \\E-mail: guenterg@desy.de}   


\twocolumn[\maketitle\abstracts{Large $\pt$ processes at HERA,
initiated by almost real and by virtual photons, provide information
on the structure of the photon. We report on the latest measurements
of dijets and large $\pt$ particle production with the H1 detector.
This includes a leading order determination of an effective virtual
photon parton density, of the gluon density of the photon, and
comparisons with models.}]

\section{What is a photon?}
\label{sec:photon}
\begin{figure}[htbp]
\begin{center}
\epsfig{file=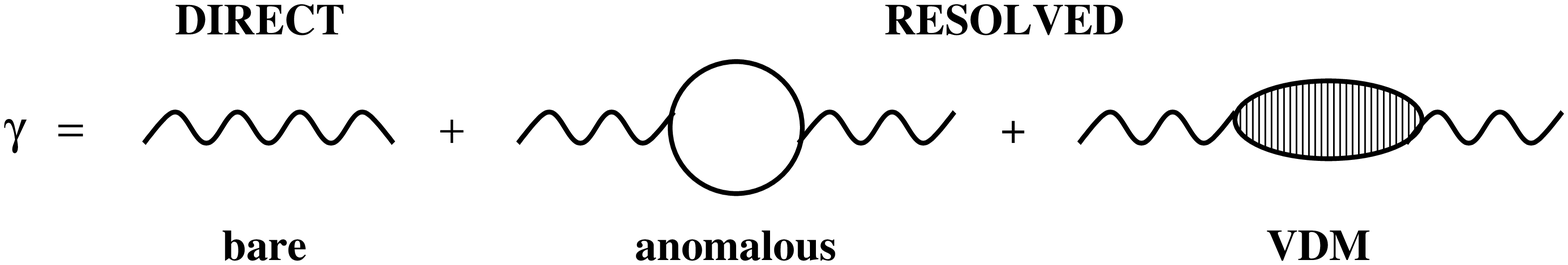,width=0.46\textwidth}
\end{center}
\caption{\label{photon}
  The dual nature of the photon and its associated nomenclature.}
\end{figure}

A photon is not just a photon and not just a hadron. In the leading
order (LO) QCD framework it has two components, a direct one, which
couples electromagnetically to the partons in the proton for example,
and a resolved one. In its resolved state it fluctuates either into an
on-shell $\qqbar$-pair forming a vector meson (VDM part) or an
off-shell $\qqbar$-pair (anomalous part), including any number of
gluons, which may interact strongly with other partons around. In
pictorial form this is expressed in Fig.~\ref{photon}.
In dijet or large $\pt$ particle production in $ep$ collisions we have
two important scales to consider, the virtuality of the photon $Q^2$
and the mean $\met$ of the hard jets or the large $\pt$ of an outgoing
charged particle. In order to gain some intuition about the effect of
the two scales, it is instructive to compare the life-times of the
$\egam$-state and the $\qqbar$-fluctuation of the photon. Making use
of the Heisenberg uncertainty relation, we find that the life-time of
the $\egam$-state falls like $1/Q^2$ (i.e. from $O(10000)$ fm to
$O(100)$ fm for $Q^2$ from $0.1$ to $100$~\gev2. The life-time of the
$\qqbar$-fluctuation is constant for fixed $\pt$ such that the photon
lives long enough do develop into a $\qqbar$ or even more complicated
state as long as $Q^2 < \ptq$. In this situation, in $ep$ collisions,
the partons in the proton are able to probe the partons in the photon
target. However with increasing $Q^2$, i.e. decreasing life-time
of the photon, the photon becomes less resolvable.

The hadronic structure of the photon has been well established both
through measurements of the real photon structure in
$\gamgam$ collisions at $\epem$ colliders (PETRA, PEP, and LEP) and
the measurements of jets in photoproduction at HERA. Extending these
measurements to the virtual photon structure is expected to provide
new insight into the QCD framework, linking deep-inelastic (DIS),
$\gamp$, and $\gamgam$ interactions.   

\setcounter{footnote}{0}
\renewcommand{\thefootnote}{\fnsymbol{footnote}}
\footnotetext[1]{Talk given on behalf of the H1 Collab., ICHEP'98,
Vancouver, Canada, July 22-30, 1998.}

\section{What is being measured at HERA?}
\label{sec:measurement}
\begin{figure}[htbp]
\begin{center}
\epsfig{file=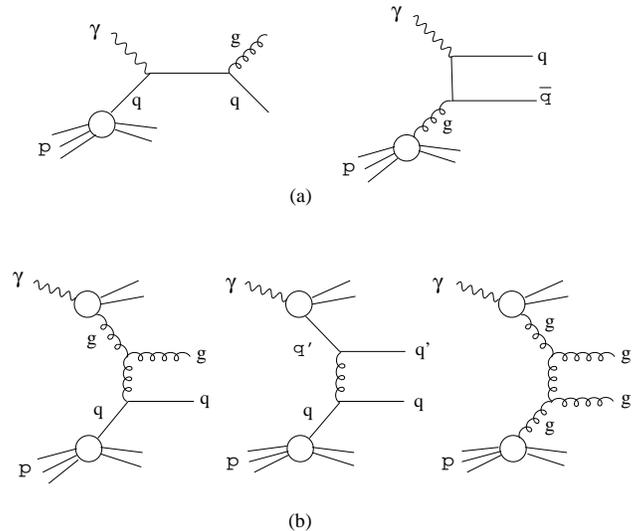,width=0.45\textwidth}
\end{center}
\caption{\label{feyn_dir_res}
  LO direct (a) and resolved processes (b)}
\end{figure}

In LO QCD we expect contributions to dijet production from the direct
processes of photon-gluon fusion and QCD-Compton and the
resolved $2 \rightarrow 2$ parton processes shown in
Fig.~\ref{feyn_dir_res}.
In next-to-leading order (NLO) direct and resolved processes can no
longer be distinguished. Experimentally, the hard scale $Q^2$ and the
fractional energy of the photon, $y$, are determined from the
measurement of the energy and angle of the scattered electron. The
variable $\xgamjets$, correlated with the fractional energy, $\xgam$,
of the parton in the photon participating in the hard scattering
process is determined from the energy and direction of the two hardest
jets and all of the hadrons in the event, using:
\begin{equation}
\xgamjets  =  \frac{\sum_{jet1}^{jet2}(\ejet - \pzjet)}{\sum_{h}(\eh - \pzh)}.
\label{equ-xgamjets}
\end{equation}
By cutting on $\xgamjets$, enriched event samples due to either direct
or resolved processes can be obtained. At the parton level $\xgamjets
= 1$ for direct processes.

The resolved $ep$ cross section can be written as:
\begin{eqnarray}
\sigepres & \sim & \frac{\fgame(y,Q^2)}{y} \nonumber\\
& & \sum_{i,j}^{N_f} \frac{\figam(\xgam,\etq,Q^2)}{\xgam}
\frac{\fjp(\xp,\etq)}{\xp} \mij2 ,
\label{equ-sigmares}
\end{eqnarray}
where the sum runs over all $2 \rightarrow 2$ hard scattering matrix
elements, folded by the density of parton $j$ in the proton and parton
$i$ in the photon, and finally folded by the density of the photon in
the electron. In the fractional energy, $\xp$, of the parton in the
proton, the measurement covers the range from about $0.01$ to $0.1$,
where the parton densities of the proton are well known.

The H1 experiment has selected events in the range $1.6 < Q^2 <
80$~\gev2 and $0.1 < y < 0.7$ in their most recent dijet
analysis~\cite{tania_dijets,tania_dijets_paper}. The jet selection was
performed in the $\gamstarp$ center of mass system (cms) using the
inclusive $k_t$ algorithm~\cite{incl_kt_algo}. Events were required to
have at least two jets satisfying the following criteria:
\begin{eqnarray}
|\etajet[1] - \etajet[2]|  <  1.0, & -3.0  <  \meta < -0.5, \nonumber\\
\metq  >  30 {\rm GeV}^2, \,{\rm and}\, &
\frac{\etjet[1]-\etjet[2]}{\etjet[1]+\etjet[2]} < 0.25,
\end{eqnarray}
where $\meta$ and $\met$ are the mean pseudorapidity
($\eta=-\ln\tan(\theta/2)$) and mean transverse energy of the two
highest $\et$ jets. The first two cuts make sure that the jets are
confined to the acceptance of the detector, where they are well
measured and that $\xgam$ is well determined. The restriction on the
difference in $\eta$ of the jets reduces the probability of
misidentifying a part of the photon or proton remnant as one of the
high $\et$ jets. The constraints are such that the highest (second
highest) $\et$ jet has $\et~\gapprox~7~(4)$~GeV. This asymmetric jet
selection will allow a comparison of the data with NLO calculations. 
With this
selection H1 obtains a sample of $\sim 12000$ dijet events for an
integrated luminosity of $6$pb$^{-1}$.

The dijet cross sections measured as a function of $\xgam$, $\metq$,
and $Q^2$ have been corrected for detector acceptance and resolution
effects in separate ranges of $Q^2$ by applying an iterative Bayesian
unfolding technique~\cite{dag_unfolding} in $\xgam$ and $\metq$. The
jet profiles and the pedestal energy outside of the jets are
reasonably well described~\cite{tania_dijets,tania_dijets_paper} by
the HERWIG~\cite{herwig} Monte Carlo (MC) used for the unfolding and
also by the RAPGAP~\cite{rapgap} MC used to estimate uncertainties due
to different models. A good description of the pedestal energy, also
referred to as soft underlying
event~\cite{tania_dijets,tania_dijets_paper} and in part caused by
soft interactions of the photon and proton remnant, is of importance
due to the steeply falling $\pt$ distribution of the jets.

For the virtual photon parton densities the Herwig MC uses the model
by Drees and Godbole (DG)~\cite{drees_godbole} for suppressing the
parton densities of real photons with increasing $Q^2$ by
interpolating smoothly between the $\ln(\ptq/\Lambda_{QCD}^2)$
behavior of real photons and the asymptotic $\ln(\ptq/Q^2)$ dependence
of the anomalous piece, i.e.:
\begin{eqnarray}
\fqgam(\xgam,\ptq,Q^2) & = & \fqgam(\xgam,\ptq,0) \, L(\ptq,Q^2,\omega) \nonumber\\
\fggam(\xgam,\ptq,Q^2) & = & \fggam(\xgam,\ptq,0) \, L^2(\ptq,Q^2,\omega),
\,\, {\rm where} \nonumber\\
L(\ptq,Q^2,\omega) & = & 
\frac{\ln\left\{(\ptq + \omega^2)/(Q^2 + \omega^2) \right\}}
{\ln\left\{(\ptq + \omega^2)/\omega^2 \right\}},
\end{eqnarray}
where $\omega^2$ is a free parameter to be chosen to describe the
data.  Another model of virtual photon parton densities is provided by
Schuler and Sj\"ostrand~\cite{pdf_sas}, consisting of a vector meson
dominance and a perturbative anomalous component with appropriate $Q^2$
evolution. 

%

\section{Triple differential cross section}
\label{sec:tripdiff}
\begin{figure}[htbp]
\begin{center}
\epsfig{file=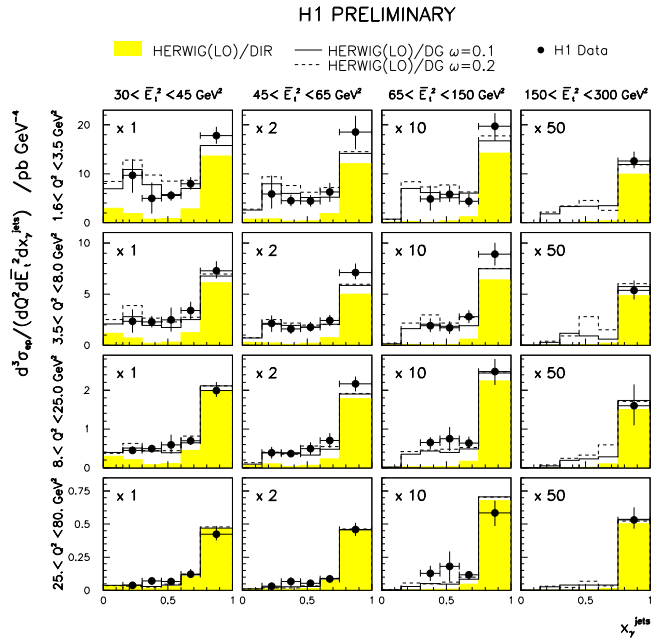,bbllx=0pt,bblly=10pt,bburx=555pt,bbury=560pt,width=0.47\textwidth,clip=}
\end{center}
\caption{\label{xgam_dijets}
  The differential dijet cross section shown as a function of
  $\xgamjets$ for different regions of $\metq$ and $Q^2$. Scale
  factors applied to the cross section are indicated. The error bars
  of the data points show the statistical and systematic errors added
  in quadrature. Also shown are model predictions (HERWIG) with $10\%$
  soft underlying event and two choices of the $Q^2$ suppression
  factor $\omega$. The direct processes as given by this model are
  indicated as shaded histogram.}
\end{figure}

\begin{figure}[htbp]
\begin{center}
\epsfig{file=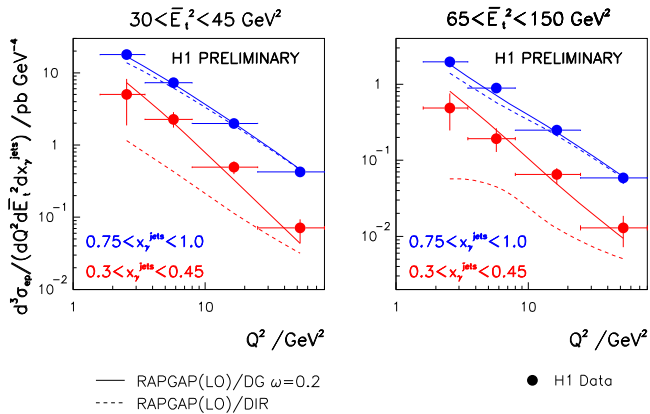,bbllx=0pt,bblly=210pt,bburx=515pt,bbury=540pt,width=\linewidth,clip=}
\end{center}
\caption{\label{q2_dijets}
  The differential dijet cross section shown as a function of $Q^2$
  for a low and a high range in jet $\metq$ and for ranges in
  $\xgamjets$ dominated by either direct or resolved processes. Also
  shown are model predictions (RAPGAP) with a $Q^2$ suppression factor
  $\omega = 0.2$. The sum of the LO direct and resolved contributions
  is indicated by the full line, the direct one by the dashed line.}
\end{figure}

The corrected triple differential cross section is shown as a function
of $\xgamjets$, $\metq$, and $Q^2$ respectively in
\cite{tania_dijets,tania_dijets_paper}. In each case the distributions
are shown for ranges of the other two variables. Here, due to lack of
space, we only show the $\xgamjets$ distributions in
Fig.~\ref{xgam_dijets}.
They can be seen to peak towards $\xgamjets=1$, where the direct
contribution is expected to dominate. For a fixed $Q^2$ or life-time
of the target photon the cross section shows a strong decrease with
increasing $\metq$ of the probing jet. Increasing $Q^2$, i.e.
decreasing the life-time of the photon, while keeping the jet $\metq$
constant, leads to a decrease of the cross section and a diminishing
relative contribution from resolved photons (for $\xgamjets \leq
0.75$). The HERWIG MC using GRV-LO~\cite{pdf_photon_grv} for the
parton densities of the real photon and the DG model with $\omega \sim
0.1$ to $0.2$ describes the data. In Fig.~\ref{q2_dijets} the $Q^2$
dependence of the target photon is shown for a low ($30 < \metq <
45$~\gev2) and a high ($65 < \metq < 150$~\gev2) range in $\metq$ of
the probing jet for a bin in $\xgamjets$ dominated by resolved ($0.3 <
\xgamjets < 0.45$) and by direct ($0.75 < \xgamjets < 1.0$) processes.
We observe a strong decrease of the cross section with $Q^2$. The
resolved dominated contribution shows an even faster decrease than the
direct dominated one. This is what we qualitatively anticipate from
the simple considerations in Sect.~\ref{sec:photon}. They are
supported quantitatively by the RAPGAP MC
as shown in Fig.~\ref{q2_dijets}, which is in good agreement with the
data. The stronger $Q^2$ suppression of the resolved contribution is
due to the additional $Q^2$ suppression by the virtual photon parton
densities.

We conclude that the observed dependence of the dijet cross section for 
$\xgamjets < 0.75$ is consistent with that predicted for a resolved 
virtual photon with parton densities evolving with $Q^2$ according to 
QCD motivated models.

\section{Effective virtual photon parton density}
\label{sec:effvpdf}
In order to determine the parton densities of the virtual photon, H1
has adapted the single effective subprocess (SES) approximation,
originally developed for use in $\ppbar$ collisions~\cite{ses} and
recently used to investigate real photon
structure~\cite{h1_real_effpdf}. This approximation exploits the fact
that the dominant contributions to the cross section comes from $2
\rightarrow 2$ scattering matrix elements which have similar kinematic
dependencies and differ mainly by their associated color factors.
Therefore they can be replaced by an effective matrix element,
$M^{SES}$, and effective parton densities for the virtual photon,
$\feffgam$, and proton, $\feffp$. The resolved $ep$ cross section of
Eq.~\ref{equ-sigmares} then becomes
\begin{equation}
\sigresses \sim \frac{\fgame(y,Q^2)}{y} 
\frac{\feffgam(\xgam,\ptq,Q^2)}{\xgam}
\frac{\feffp(\xp,\ptq)}{\xp}\, |M^{SES}|^2,
\label{equ-sigresses}
\end{equation}
where the effective parton densities are given by:
\begin{equation}
\feffgamp = \sum^{N_f} (\fqgamp + \fqbargamp) + \frac{9}{4}\fggamp
\end{equation}
and the sum runs over all quark flavors.

For the extraction of the effective parton density only data with $0.2
< \xgamjets < 0.7$ and $\etjetq > Q^2$, i.e. the condition for
resolving the partons in the photon, are used for a second unfolding
to correct the dijet cross section to the LO diparton cross section.
This unfolding tries to correct for hadronisation effects and initial
and final state QCD radiation.  The systematic
error~\cite{tania_dijets,tania_dijets_paper} includes those associated
with the determination of the triple differential cross section and
additional errors arising in the second unfolding and amounts to an
average error of $\sim 40\%$.

\begin{figure}[t!]
\begin{center}
\epsfig{file=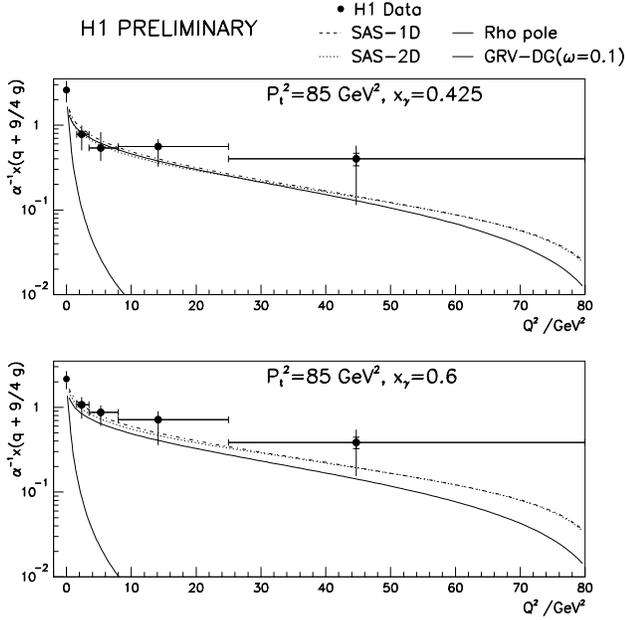,bbllx=15pt,bblly=20pt,bburx=545pt,bbury=550pt,width=0.45\textwidth,clip=}
\end{center}
\caption{\label{effpdf_q2}
  The LO effective parton density of the photon, divided by the
  fine-structure constant $\alpha$, as a function of photon virtuality
  $Q^2$ for $\ptq=85$~\gev2 and two values for $\xgam$. The error bars
  of the data points indicate the statistical and systematic errors
  added in quadrature. Also shown are predictions from the DG model
  with $\omega = 0.1$, using GRV-LO parton densities for real photons
  and the SAS-1D (dashed line) and SAS-2D (dot-dashed line)
  parametrisations. The steeply falling solid curve shows the
  photoproduction data point extrapolated by a $\rho$-pole factor.}
\end{figure}

The resulting effective parton densities, divided by the
fine-structure constant $\alpha$, are shown in
\cite{tania_dijets,tania_dijets_paper} as a function of $\xgam$,
$\ptq$, and $Q^2$ respectively. In $\xgam$ they show a small rise
towards high $\xgam$, in $\ptq$ they are consistent with a scaling
behavior $\sim \ln\ptq$, and in $Q^2$ they are suppressed with
increasing virtuality of the photon as predicted by QCD. They are
reasonably well described by GRV-LO and the Drees and Godbole model
with $\omega \sim 0.1$ and the parametrisations SAS-1D and SAS-2D of
the model by Schuler and Sj\"ostrand. In Fig.~\ref{effpdf_q2} we show
only the results as a function of $Q^2$ for a fixed scale
$\ptq=85$~\gev2 of the probing jet and two different bins in $\xgam$.
Also shown are two photoproduction data points from
H1~\cite{h1_real_effpdf}. The $Q^2$ evolution of the data are
compared to the two models mentioned above and to a single $\rho$-pole
suppression factor, $(m^2_{\rho}/(m^2_{\rho} + Q^2))^2$, characteristic
of a simple VMD model. The $\rho$-pole factor clearly underestimates
the data, while the logarithmic suppression, as expected from QCD with
decreasing life-time of the photon, is in agreement with the
observation.
     
\section{Gluon density of the photon}
\label{sec:glue}
Measurements of the photon structure in $\epem$ interactions are
sensitive to the quark structure of the photon and only indirectly
through the QCD evolution to the gluon structure. The data so far have
not been precise enough to allow a determination of the gluon density
of the photon. Recently, studies of photo-produced dijets and high
$\pt$ charged particles at HERA have shown that these data are
sensitive to both the quark and gluon content of the photon. A first
measurement by H1 of the LO gluon density~\cite{h193_dijets} showed
that it is not large at high $\xgam$, in contrast to the
LAC3~\cite{pdf_lac} parametrisation, and does not have a steep rise
towards low $\xgam$ as predicted by the LAC1~\cite{pdf_lac}
parametrisation. The minimum $\xgamjets$ value which was reached in
this measurement was $0.04$. It is of interest to reach $\xgam$ values
as low as possible, since most suggested gluon densities are predicted
to rise. Experimentally it is rather difficult to reach low $\xgam$.
Because of the relation $\xgam \sim \et \exp^{-\eta}/(2E_{\gamma})$
for $Q^2 \sim 0$~\gev2, low $\xgam$ implies small $\et$ (in conflict
with a good correlation with the hard dipartons), large $\eta$ (soft
underlying event and detector acceptance), and large $E_{\gamma}$
(decreasing event rate).

H1 has contributed two different analyses, one on 
dijets~\cite{oliver_dijets} which is still in progress and one on 
high $\pt$ particles~\cite{highpt} which has become 
final~\cite{highpt_paper} after the conference. The latter analysis 
does not suffer from the energy scale uncertainty of the calorimeter 
and is less sensitive to the soft multiple interactions compared to 
the dijet analysis, but has the drawback of stronger sensitivity to 
uncertainties in the fragmentation.

\begin{figure}[htbp]
\begin{center}
\epsfig{file=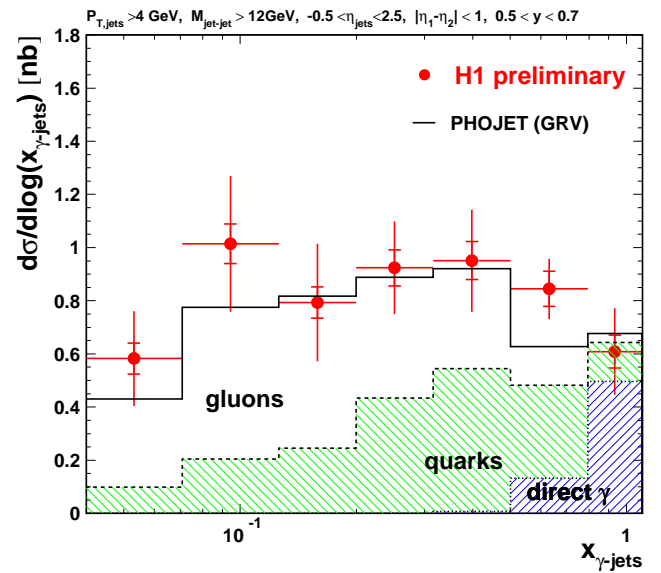,bbllx=30pt,bblly=170pt,bburx=540pt,bbury=630pt,width=\linewidth,clip=}
\end{center}
\caption{\label{oliver_xgam}
  The $ep$ dijet cross section for $Q^2 < 0.01$~\gev2 as a function of
  $\xgamjets$. The total error bars on the data points reflect the
  statistical and systematic errors added in quadrature. The data are
  compared to a LO model (PHOJET) using the GRV-LO parton densities
  for the photon and the proton. The contributions from direct and
  resolved, quark or gluon initiated, processes are indicated.}
\end{figure}

Both analyses require the scattered electron to be tagged, which
restricts the photon virtuality to $Q^2 < 0.01$~\gev2. The energy
fraction $y$ of the radiated photon is required to be in the range
$0.5$ ($0.3$ for the high $\pt$ analysis) $< y < 0.7$. In the dijet
analysis at least two jets have to be found using a cone
algorithm~\cite{cdf_cone} with a cone radius of $0.7$. Additional
requirements on the jets are: $\ptjet > 4$~GeV, $M_{jet1,jet2} >
12$~GeV, $-0.5 < \etajet < 2.5$ and $|\etajet[1] - \etajet[2]| < 1$ in
the HERA system.  Comparisons of data and MC can be found
in~\cite{oliver_dijets}. The corrected dijet cross section as a
function of $\xgamjets$, after unfolding~\cite{dag_unfolding}, is
shown in Fig.~\ref{oliver_xgam}.
It is compared to LO predictions by the PHOJET~\cite{phojet} MC using
the GRV-LO parton densities for the photon and the proton. The
transverse energy $\et$ of the jets is used for both the
renormalisation and factorisation scale. Also indicated are the
different contributions from the direct and resolved processes; the
latter are split into contributions initiated from either a quark or a
gluon on the photon side. It is clear from the figure that there is a
sizeable contribution from gluons and that the data are precise enough
to constrain the gluon density with a precision of $\approx 30\%$.

\begin{figure}[htbp]
\begin{center}
\epsfig{file=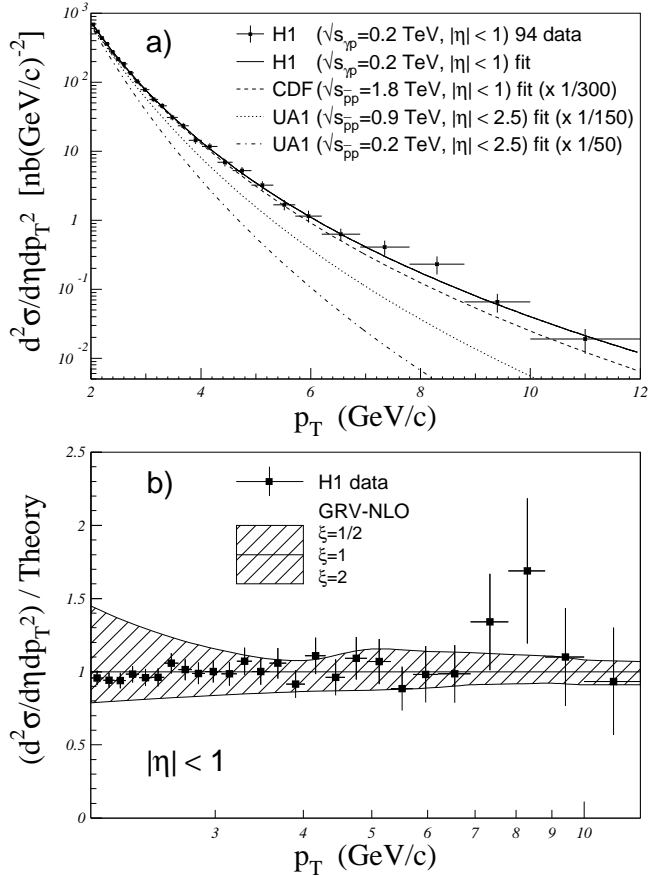,bbllx=45pt,bblly=90pt,bburx=550pt,bbury=775pt,width=\linewidth,clip=}
\end{center}
\caption{\label{highpt_fit_nlo}
  a) The inclusive $\gamp$ cross section for charged particles as a
  function of $\pt$. The error bars on the data points denote the
  statistical and systematic errors added in quadrature. The curves
  indicate power-law fits to the H1, CDF, and UA1 data; the latter
  have been normalised to the H1 data at $\pt=2$~GeV. b) The ratio of
  data to an NLO QCD calculation with scale $\xi \ptq$.}
\end{figure}

In the high $\pt$ charged particle analysis the further requirements 
are: tracks which have a $\pt > 2$~GeV and $|\eta| < 1$ in the HERA 
laboratory frame. In Fig.~\ref{highpt_fit_nlo}a the corrected $\pt$ 
distribution and fits to the H1  and $\ppbar$ data are shown.
A QCD inspired power-law expression of the form $A(1+\pt/p_{\rm T0})^{-n}$
was fit. The fit gives $A=5.44 \pm 0.66$~mb and $n=7.03 \pm
0.21$~(stat.+syst.) and describes the data well over the whole range
in $\pt$. Similar fits to $\ppbar$ data from UA1~\cite{ua1_pt} and
CDF~\cite{cdf_pt} have been normalised to the $\gamp$ cross section at
$\pt=2$~GeV. The high $\pt$ tail in the $\gamp$ data is clearly larger
than in $\ppbar$ collisions at similar cms energies. This can be
understood as being due to extra contributions in $\gamp$, namely the
direct and the point-like resolved component. Similar conclusions have
been obtained for $\gamgam$ collisions by
OPAL~\cite{soeldner-rembold}.

\begin{figure}[htbp]
\begin{center}
\epsfig{file=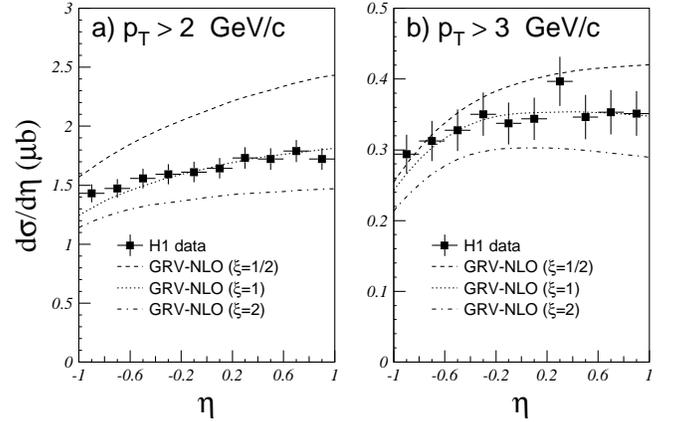,bbllx=60pt,bblly=270pt,bburx=530pt,bbury=590pt,width=\linewidth,clip=}
\end{center}
\caption{\label{highpt_eta_nlo}
  The inclusive $\gamp$ cross section as a function of $\eta$ for
  charged particles with a) $\pt > 2$~GeV and b) $\pt > 3$~GeV in
  comparison with an NLO QCD calculation with scale $\xi \ptq$. The
  error bars indicate the statistical and systematic errors added in
  quadrature.}
\end{figure}

\begin{figure}[htbp]
\begin{center}
\epsfig{file=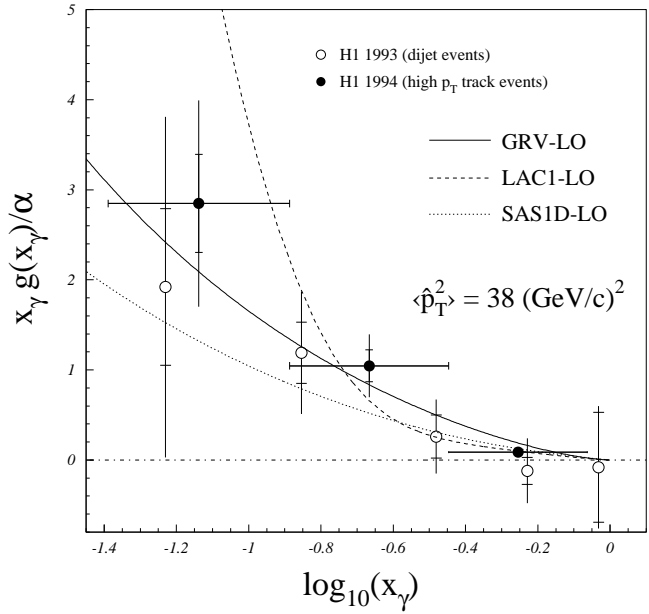,bbllx=35pt,bblly=20pt,bburx=530pt,bbury=490pt,width=\linewidth,clip=}
\end{center}
\caption{\label{highpt_xg}
  The LO gluon density in the photon from charged high $\pt$ tracks
  (full circles) and from dijets (open circles) with a mean scale
  $\ptq = 38$~\gev2 and $\ptq= 75$~\gev2 respectively. The full error
  bars give the statistical and systematic errors added in quadrature.
  The lines show the predictions of three different parametrisations.}
\end{figure}

The ratio of data to theory, as given by an NLO calculation including 
direct and resolved contributions~\cite{kniel_kramer}, and using as 
scales $\mu^2_{\gamma} = \mu^2_{p} = \mu^2_{\rm had} = \xi \pthadq$ 
with $\xi = 1/2, 1,$ and $2$, is shown in Fig.~\ref{highpt_fit_nlo}b.
The pseudorapidity distributions for data and the NLO calculation 
for $\pt > 2$ and $\pt > 3$~GeV are given in Fig.~\ref{highpt_eta_nlo}. 
The data are well described by NLO, in particular for $\xi$ close to 
$1$. However, one finds that the NLO prediction is rather sensitive to 
the choice of scale ($\xi$). This effect is smaller for larger $\pt$.
For each event with at least one charged particle with $\pt >
2.6$~GeV, the variable $\xgamrec = \sum \pt \exp(-\eta)/E_{\gamma}$,
where the sum runs over all tracks with $\pt > 2$~GeV, is calculated.
It shows a good correlation~\cite{highpt_paper} to the true $\xgam$.
To obtain $\xgam$ an unfolding procedure~\cite{blobel_unfolding} was
used.  Then the LO gluon density was unfolded allowing for an
uncertainty in the quark densities of $\sim 15\%$ as given by three
different parametrisations~\cite{pdf_photon_grv,pdf_sas,pdf_lac} and
using different hadronisation models. The resulting gluon density
(only available after the conference) is shown in Fig.~\ref{highpt_xg}
as a function of $\xgam$ and compared to an older extraction using
dijets~\cite{h193_dijets} and to three different parametrisations.
The mean $\ptq$ of the hard scattering process for this data sample is
$38$~\gev2 according to MC, which was used as the scale for the
comparison with three parton density parametrisations,
GRV-LO~\cite{pdf_photon_grv}, SAS-1D-LO~\cite{pdf_sas}, and
LAC1-LO~\cite{pdf_lac}. The mean $\ptq$ for the dijet sample is
$75$~\gev2. The results confirm that the contribution of the gluon to
the photon structure is significant. The gluon density rises
with decreasing $\xgam$ and is best described by GRV-LO.

\section{Summary}
At HERA, with dijet and high $\pt$ particle production, we can probe
the virtual photon structure over a large range in the scale $\sim
\ptq$ and in the life-time of the photon $\sim 1/Q^2$.

In the measured dijet cross section as well as the extracted effective
virtual photon parton density, we observe the expected logarithmic
suppression with $Q^2$, as the life-time of the photon decreases, of
the resolved photon contribution.

Using two complementary methods, high $\pt$ charged particles and
dijets in almost real photoproduction, the leading order gluon density
of the photon was determined and found to be rising with decreasing
$\xgam$.

So far we have only scratched the surface in that we have found
consistency with rather global expectations and in that we have
learned how to do leading order determinations of parton
densities. There is a lot more to do, stay tuned for 1999.

\section*{Acknowledgements}
I want to thank my colleagues in H1, who helped in preparing the
presentation, and Martin Erdmann and Steve Maxfield for reading this
manuscript.

\end{document}